# Investigation of the specific plasma potential oscillations with geodesic acoustic mode frequencies by heavy ion beam probing

A V Melnikov[1], L G Eliseev[1], A V Gudozhnik[1], S E Lysenko[1], V A Mavrin[1], S V Perfilov[1], M V Ufimtsev[2], L I Krupnik[3], L G Zimeleva[1]

[1] Nuclear Fusion Institute, RRC "Kurchatov Institute", 123182, Moscow, Russia,
[2] Department of Computational Mathematics and Cybernetics, Moscow State University, Moscow, Russia
[3] Institute of plasma physics, NSC “Kharkov Institute of Physics and Technology”, Kharkov, Ukraine
E-mail: melnik@nfi.kiae.ru

**Abstract**

Investigation of the specific oscillations with frequencies 15-30 kHz on the T-10 tokamak ($R$= 150cm, $a$ = 30 cm) with Heavy Ion Beam Probe (HIBP) diagnostic was conducted in regimes with off-axis ECRH ($B$ = 2.33 T, $I_p$= 220 kA, $r_{ECRH}$=12 cm). Previous experiments in OH regimes have shown that “20 kHz” modes are mainly the potential fluctuations. These oscillations are seen on the signals of HIBP, Langmuir probes and reflectometry. They should cause the fluctuations of the poloidal rotation, i.e. the torsional plasma oscillations with $m$=0, called as the zonal flows. The HIBP sample volume was localized at $r$ = 24 - 29 cm. It was observed that both in OH and ECRH regimes, the power spectrum of potential oscillations has a form of solitary quasi-monochromatic peak with the contrast range of 3-5. The frequency of “20 kHz” mode is varied in the region of observation; it diminishes to the plasma edge from 20 kHz at 22-24 cm till 13-15 kHz at 28-29 cm. After ECRH switch-on, the frequency increases, correlating with growth of the electron temperature $T_e$, measured by 2-nd ECE harmonic on the nearest chord 24 cm. Analysis have shown that frequency of the “20 kHz” mode is varied with local $T_e$: $\nu_{20}$~$T_e^{1/2}$ that is similar to a theoretical scaling for Geodesic Acoustic Modes (GAM): $\nu_{GAM}$~$c_s/R$~$T_e^{1/2}$, where $c_s$ is a sound speed. The absolute frequencies are close to GAM values within a factor of unity (1.12-1.5).

## 1. Introduction

The turbulence driven poloidal flows, the zonal flows, are thought to be a mechanism affected the turbulence and transport in magnetically confined plasmas. One of manifestation of the zonal flows is the oscillations in plasma radial electric field and, therefore, in plasma potential. Geodesic Acoustic Mode (GAM) is the high-frequency class of the zonal flows.

Specific oscillations of the plasma electric potential with frequencies near 20 kHz, called as “20 kHz” mode (15 – 30 kHz) have been discovered in the TEXT tokamak by Heavy Ion Beam Probe (HIBP) diagnostic in 1993 [1, 2]. During 2000-2002, the DIII-D team observed a peak on the spectrum of poloidal rotation rates in the range ν=13-16 kHz. They proposed to link it with the GAM, which is characterized by the square root dependence on the local electron temperature $T_e$ [3, 4]. In both cases the spectra look like the strongly dominated sharp monochromatic peak on the noisy background. On the T-10 tokamak, the typical peak was observed at the appropriate frequencies on the signals of HIBP, Langmuir probes and reflectometry [5]. HIBP is used for measurements of the plasma electric potential on many magnetic fusion devices. Basic principles of HIBP diagnostic are reviewed in [6].

Here we discuss the dynamics of the specific oscillations of the potential with frequencies 15-30 kHz measured by HIBP in the regimes with Ohmic and ECR heating on the T-10 tokamak. The main parameters of discharge are: $R$=150 cm, $a$ =30 cm, $B_0$ = 19.5-2.33 T, $I_p$=180-220 kA.

On T-10, HIBP measurements were performed in two main operating modes: former uses the fixed position of sample volume, when the radial dependencies obtained from shot-to shot; the latter uses the radial scan, when the sample volume moves over the plasma cross-section with repetition rate ~50 kHz, producing a series of profiles during a single shot.

## 2. Data processing

The temporal evolution of “20 kHz” on the spectrum of HIBP signal is investigated. The signal of potential is obtained as the relative difference of currents onto the upper and lower detector plates of the HIBP analyzer:

$$\varphi \sim \delta i = (i_{UP} - i_{DOWN})/ (i_{UP} + i_{DOWN}). \qquad (1)$$

The sampling time is 11 μs that corresponds to $\nu_{Nyquist}$ = 46 kHz; the transmission bandwidth of the electronics is 50 kHz. In several measurements the sampling time was diminished to 4 μs. For the fixed position of the sample volume, potential spectra are calculated using 1024 time points (11 ms/spectrum). The width of Hanning window is 128 points. Totally it was 15 Hanning windows. The ultimate spectrum was obtained by averaging over 15 spectra from each window. For the radial scans, spectra are calculated using 64 points (0.6 ms/spectrum). Figure 1 shows that the hardware and plasma load noises are ± 5 ADC (analog-digital converter) counts, while the amplitude of signal oscillations is ~30 ADC counts, i.e. confidently detected.

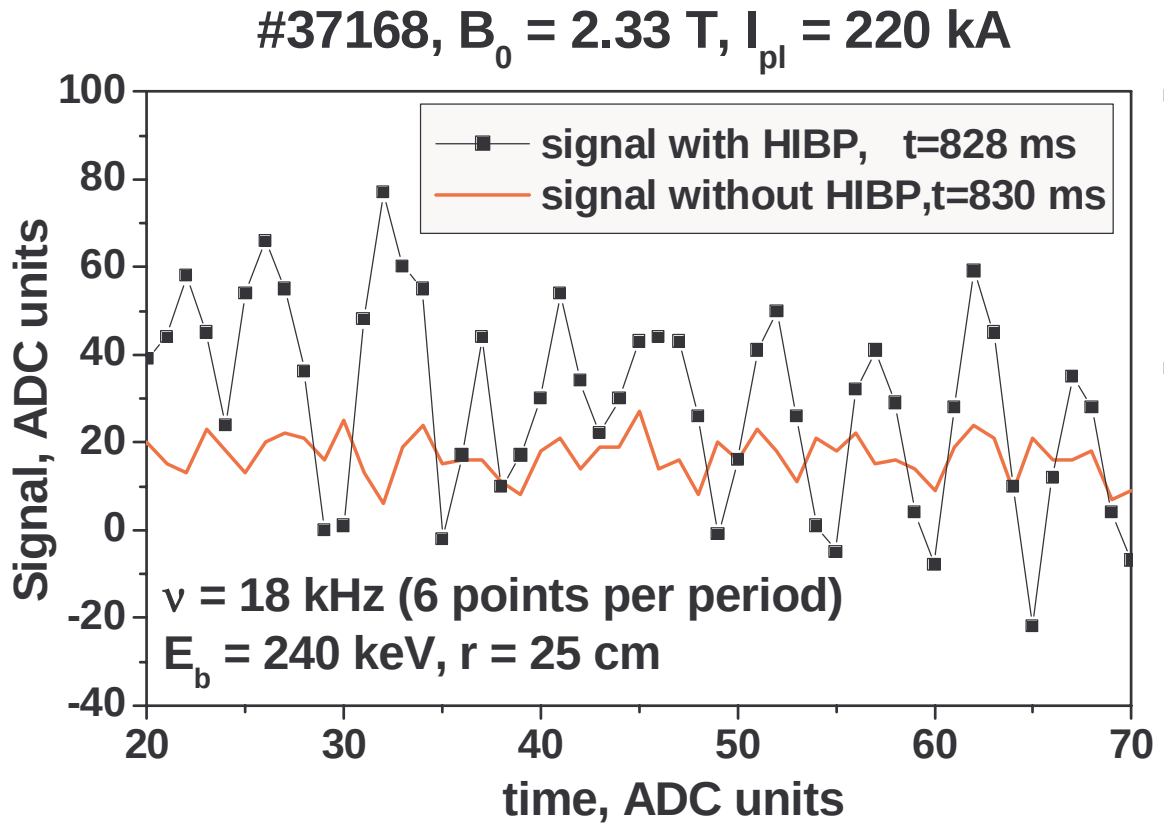

**Figure 1.** The typical time trace of the current at the detector plate with (black line with squares) and without (red line) HIBP.

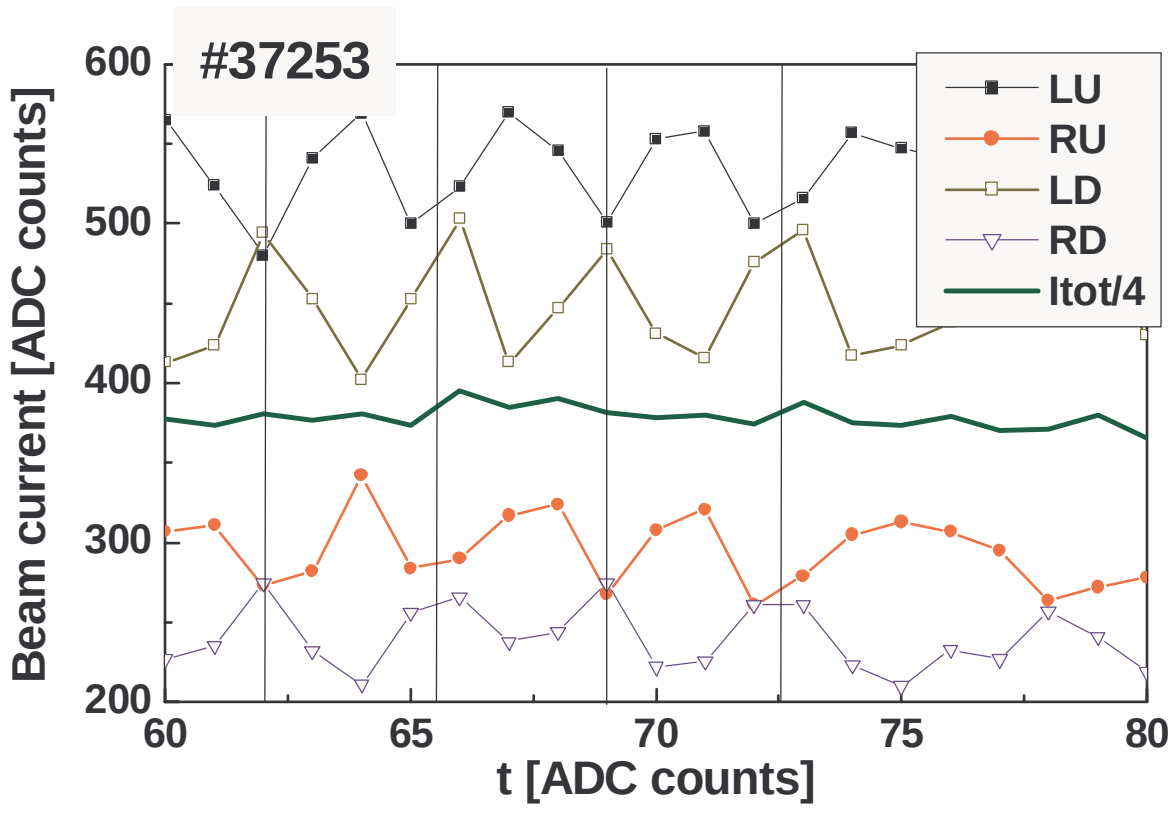

**Figure 2**. Typical forms of raw signals on the detector plates (curves are shifted for visualization).

Figure 2 presents typical forms of raw signals on the detector plates. We see that the amplitude of “20 kHz” mode is 40 ADC counts on the background of 400 ones, i.e. about 4 points per period of “20 kHz” mode. On the total current the “20 kHz” mode is not seen.

In the case of four plates the oscillations of plasma potential are calculated as follows:

$$\Delta\varphi = 2U_{an} F \delta i$$
$$\delta i = (i_{LU} + i_{RU} - i_{LD} - i_{RD}) / \Sigma i \qquad (2)$$

where $U_{an}$ is the analyzing voltage, $F$ is the dynamic factor of energy analyzer determined by special calibration o, $i$ are currents on the left (index $L$), right ($R$), upper ($U$) and lower ($D$) plates, $\Sigma i$ is the total current at all plates.

Figure 3 shows that the amplitude of "20 kHz" mode is up to A20~100 V.

Data processing for the radial scan is the same as for fixed point. For the radial scans, spectra are calculated using 64 points (0.6 ms/spectrum), which is much less than at the fixed point. To improve statistics, we match the rate of scan and the width of the window. Also we averaged several (up to 5) spectra obtained during steady state phases of discharge (OH or ECRH).

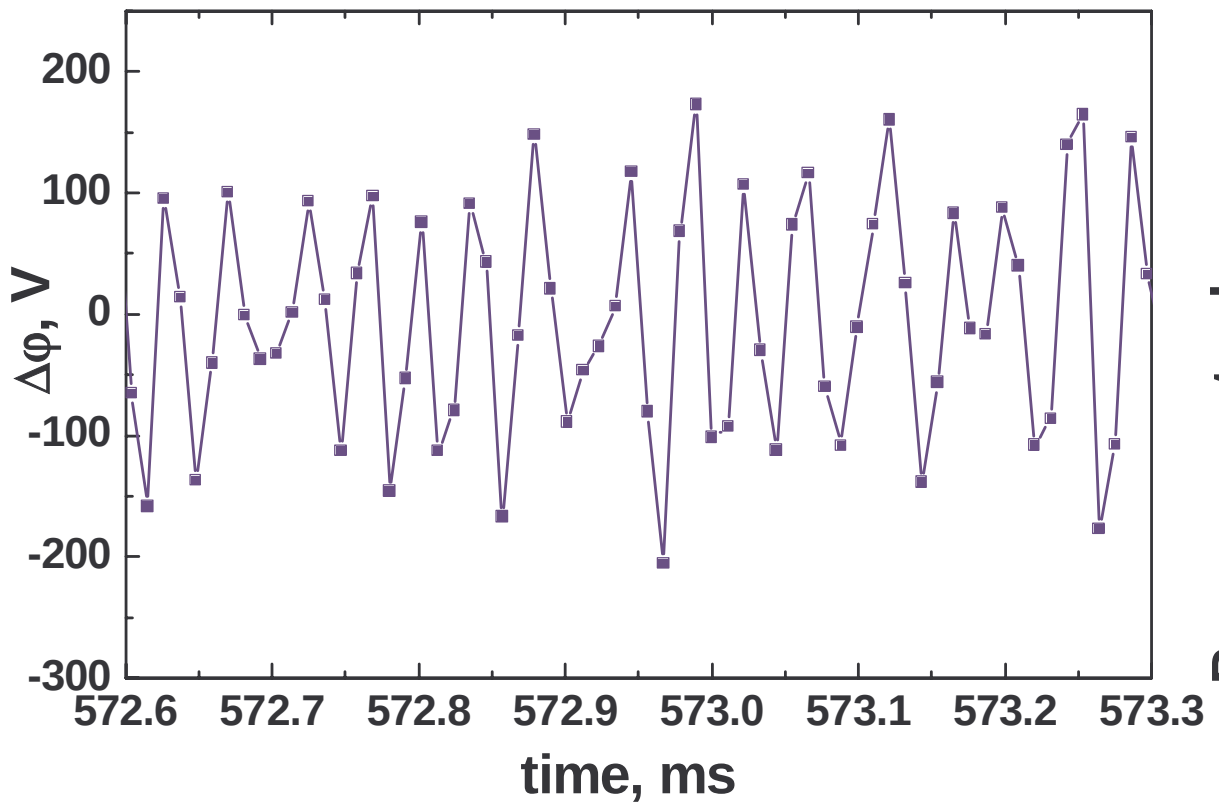

**Figure 3.** Signal of plasma potential oscillations.

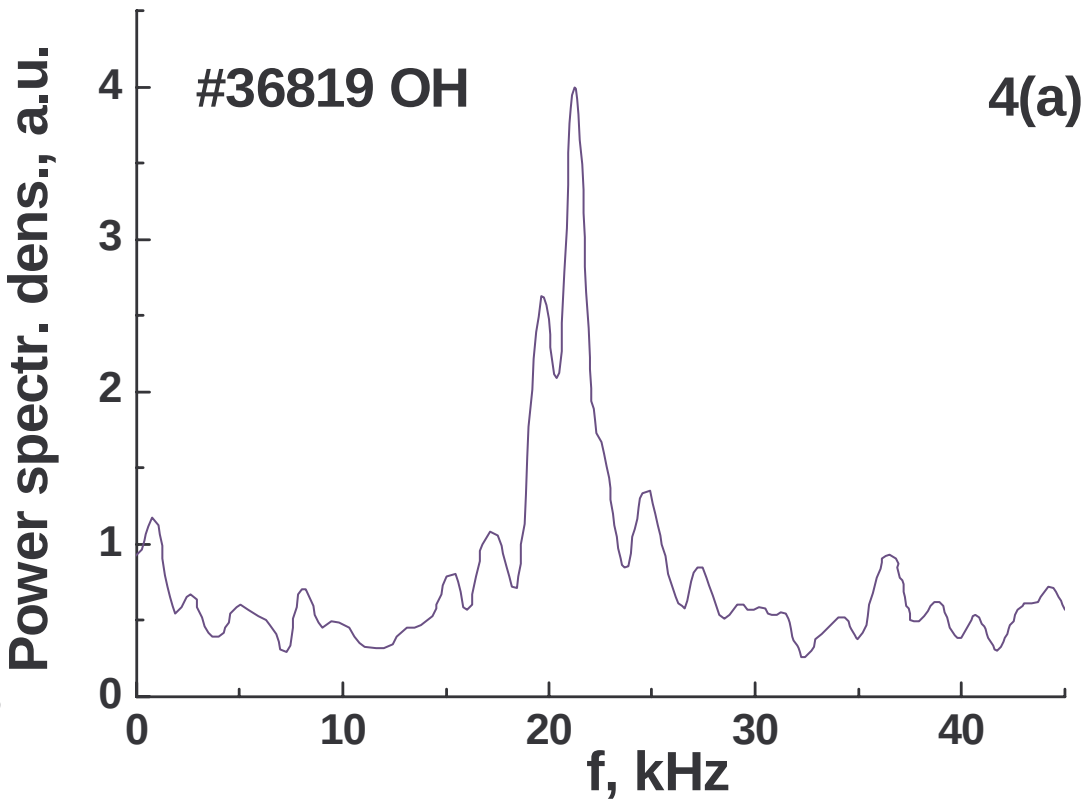

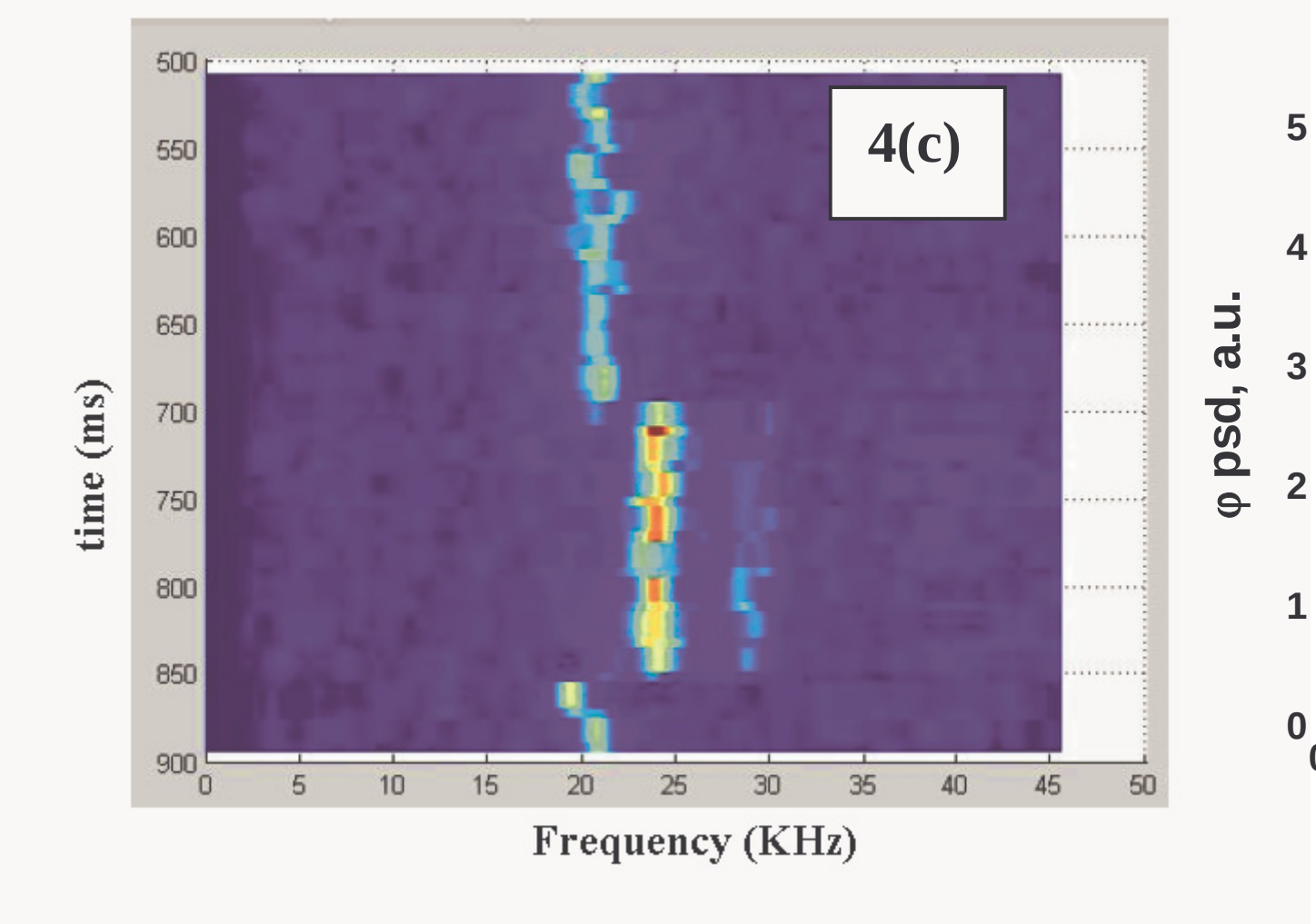

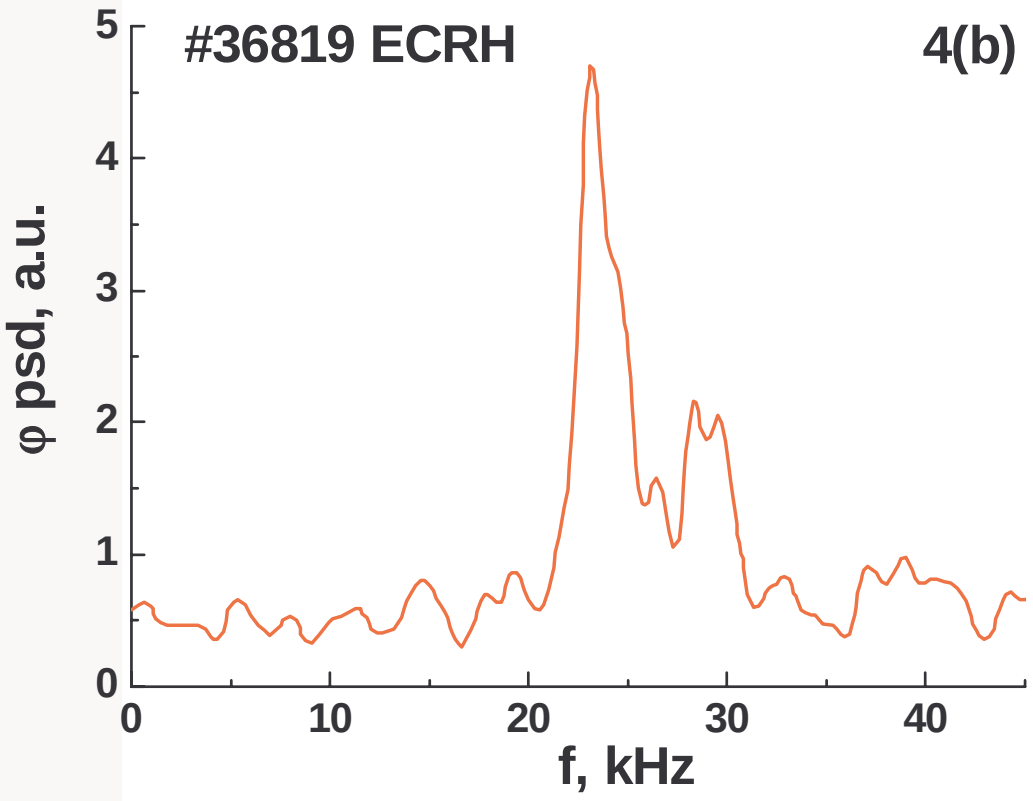

**Figure 4.** The evolution of potential spectrum: at Ohmic (a) and ECR heating (b) phases of shot #36819 ($B_0$ = 2.05 T, $I_p$ = 270 kA) at fixed sample volume, $r$ = 17 cm, $E_b$ = 240 keV. The characteristic frequency evolves with time from 20 to 24 kHz (c).

## 3. Measurements at the fixed sample volume

The typical power spectra of potential oscillations for Ohmic and ECRH phases of shot #36819 are shown in Fig. 4. Measurements in the fixed position of the sample volume, $r$ = 17 cm show that during ECRH, (700 ms $<t<$850 ms) the frequency of oscillations increases. Together with the main peak (~24 kHz), the satellite peak with higher frequency (~28 kHz) often appears. As a rule, it has smaller amplitude with respect to the main one, so we discuss only the evolution of main peak with lower frequency.

Comparison of "20 kHz" mode evolution measured at $r$ = 25 cm with the electron temperature measured by ECE at the chord 24 cm is shown in Fig. 5. Here we also see the frequency increase during ECRH, but this increase is not so pronounced as in Fig. 4.

Figure 6 shows variation of “20 kHz” mode during ECRH switch-on for the shot shown in Fig. 5. We see that the characteristic time of frequency variation coincides with characteristic time of local electron temperature variation, ~ 7 ms.

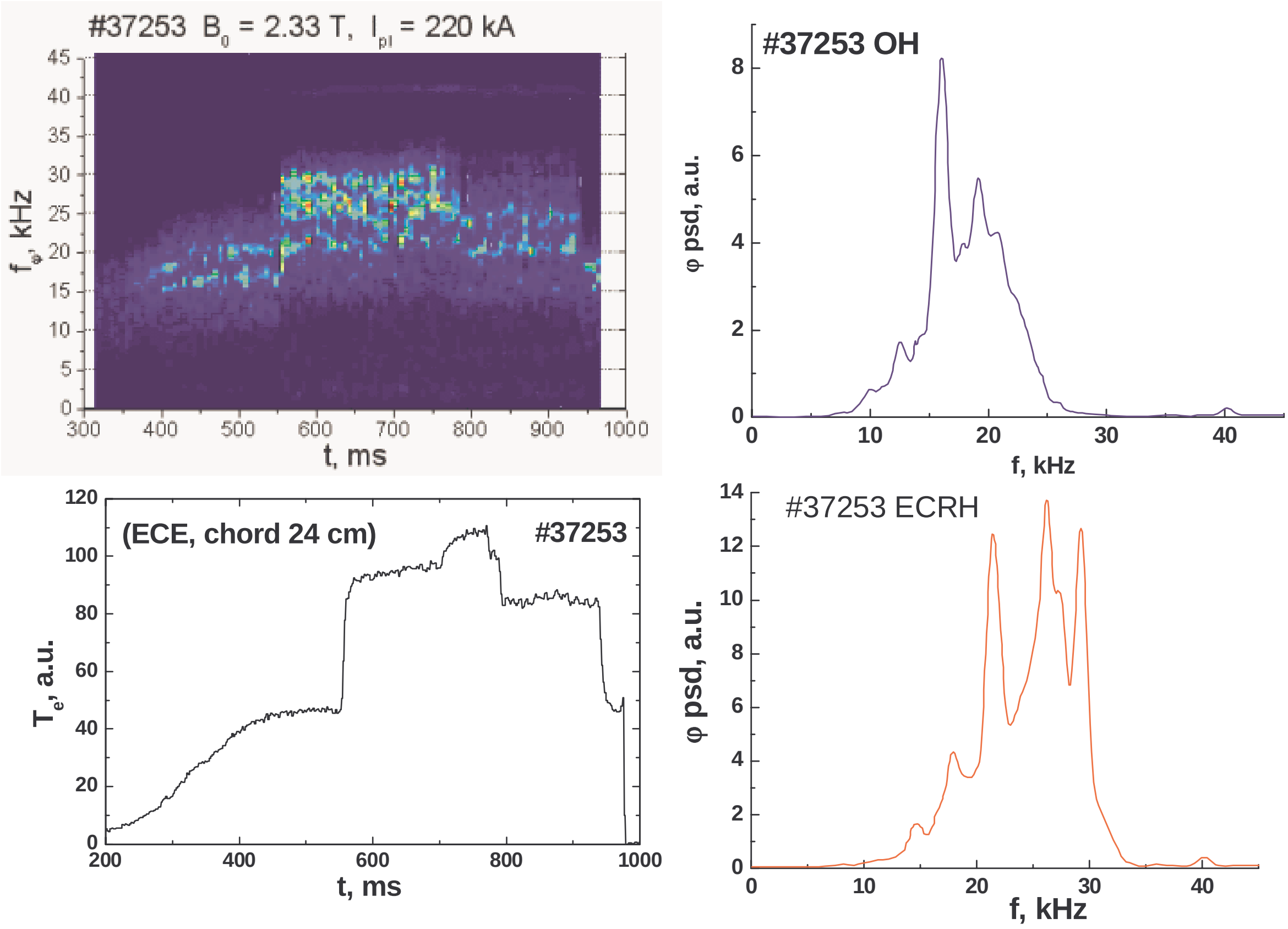

**Figure 5.** Typical spectra for OH and ECRH phases of discharge measured at $r$=25 cm. The mode frequency is varied together with temperature $T_e$. #37253, $B_0$ = 2.33 T, $I_p$ = 220 kA $E_b$= 240 keV.

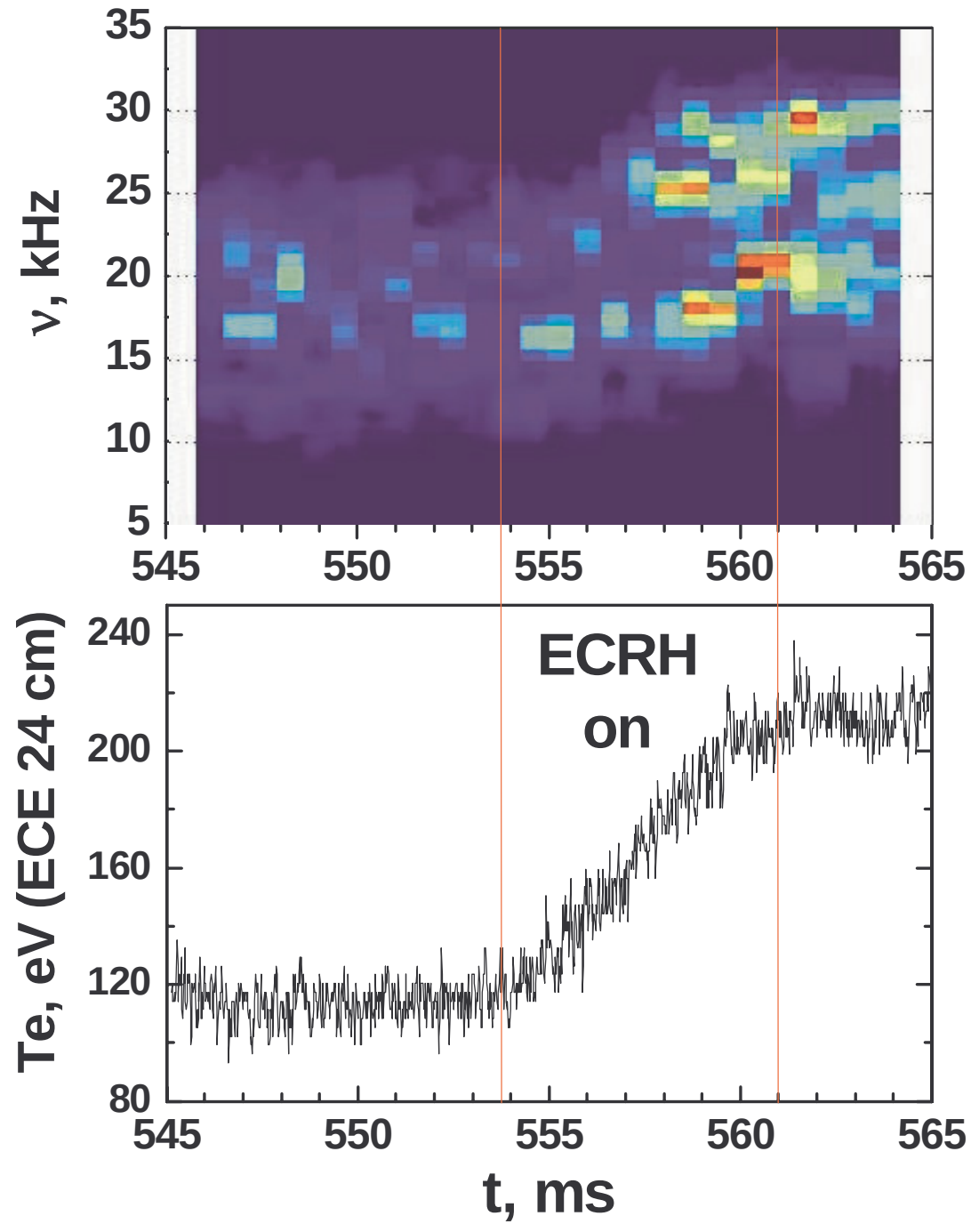

**Figure 6.** Evolution of spectrum measured at $r$ = 25 cm during ECRH switch-on. The frequency of the peak increases together with growth of electron temperature.

Figure 7 shows that in the outward region the character of evolution is the same as in the core, but the absolute value of the frequency of dominant peak is less. It evolves from 9 kHz in OH phase to 13 kHz in ECRH phase. Measurements with more rapid sampling, $\nu_N$ = 117 kHz, have shown that the "20 kHz" peak remains dominant in the extended frequency domain.

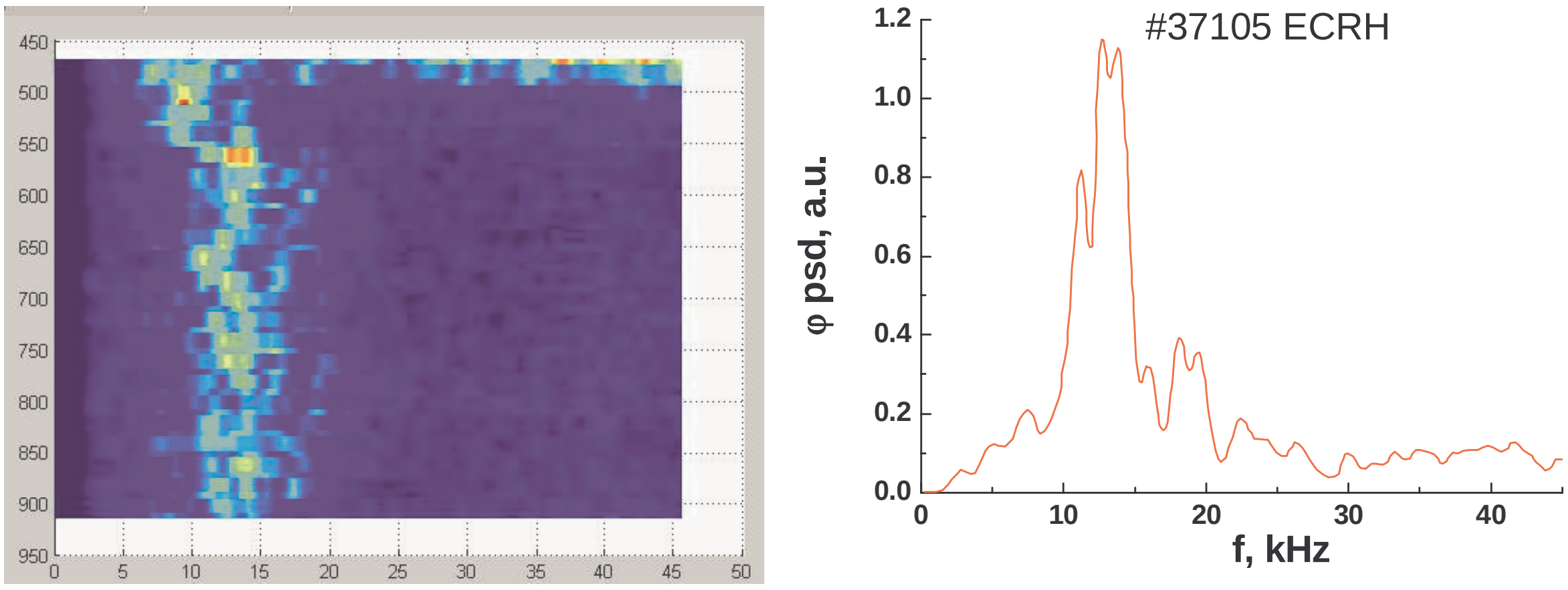

Figure 7. Spectrum variation at the outward region (**r**/*a* > 0.9). #37105

### 4. Measurements with radial scan

Figure 8 shows spectra obtained by the radial scan in the range 22-28 cm in OH and ECRH phases. We see that the peak frequency monotonically decreases with radius. The amplitude varies non-monotonically that may evidence existence of some radial structures, i.e. magnetic islands.

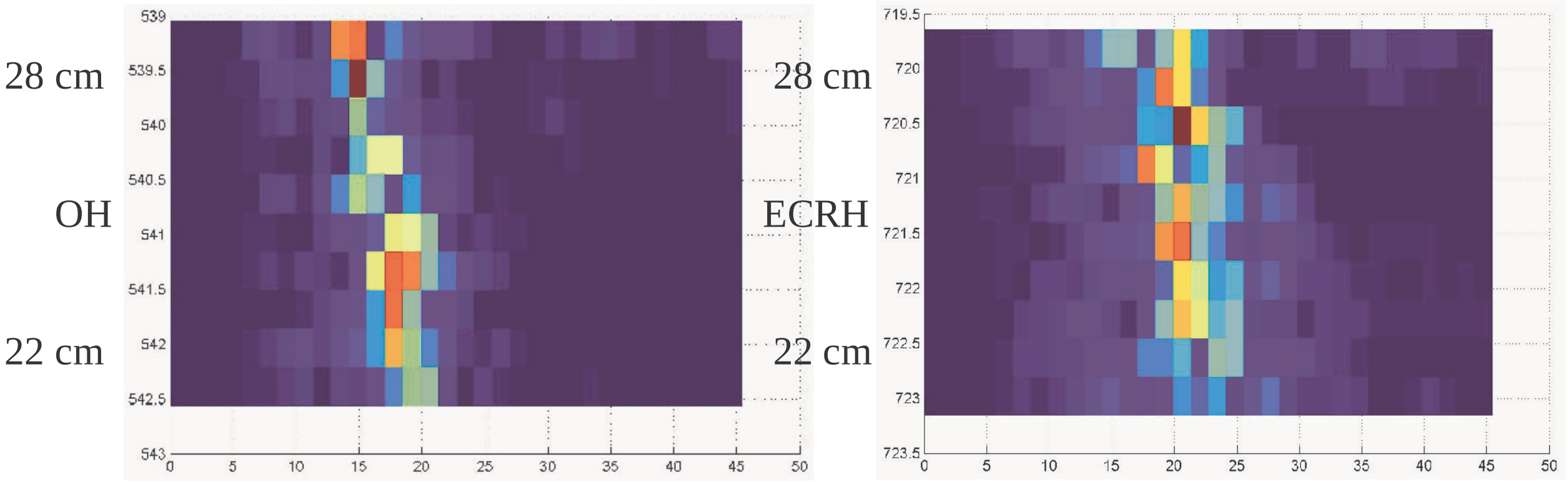

Fig. 8. Variation of spectrum with radial scan; #37242, $B_0$ = 2.33 T, $I_p$ = 220 kA, $E_b$ = 240 keV.

### 5. Discussion

The temperature dependence during one typical shot is presented in Fig. 9. The presented data were obtained during the current rump-up, steady state OH and ECRH phases. Figure 10 shows data obtained from various shots with on-axis and off-axis ECRH at $B$ = 2.33, 20.5, 19.5 T. Points measured in OH and ECRH steady states of each shot are connected.

Analysis of the experimental data has shown that the frequency of “20 kHz” mode is a rather weak function of electron temperature, it is close to be proportional to the square root of $T_e$: $\nu \sim T_e^{1/2}$ in the observed area. Such dependence is an indication of the nature of the sound waves. It allows us to interpret these potential oscillations as the Geodesic Acoustic Mode (GAM), a high-frequency type of zonal flows [2,3].

For more accurate comparison with GAM frequency, we used more precise two-fluids expression for GAM frequency, taking into account the ion temperature:

$$\nu_{GAM} = \frac{1}{2\pi R}\sqrt{(T_e + \tfrac{5}{3}T_i)/M_i} \qquad (3)$$

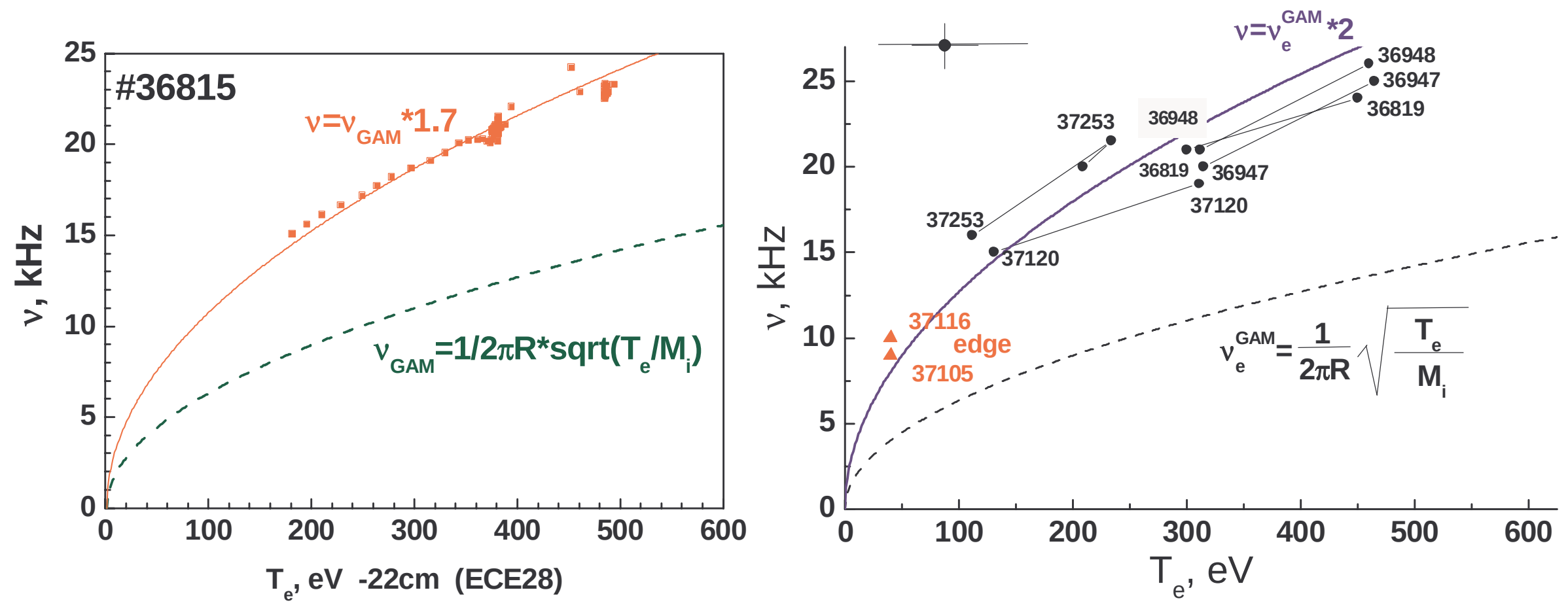

**Figure 9.** The “20 kHz” mode frequency as a function of the local electron temperature. Measurements in different phases of single shot (left) and in many shots (right). The dependence for Geodesic Acoustic Mode is also shown.

The ion temperature was calculated by the transport code ASTRA assuming experimental $T_e$ profile, the doubled neoclassical diffusivity and classical electron-ion equipartition. Figure 10 shows that the experimental frequency is close to theoretical, but higher than the latter one in a factor of 1.12.

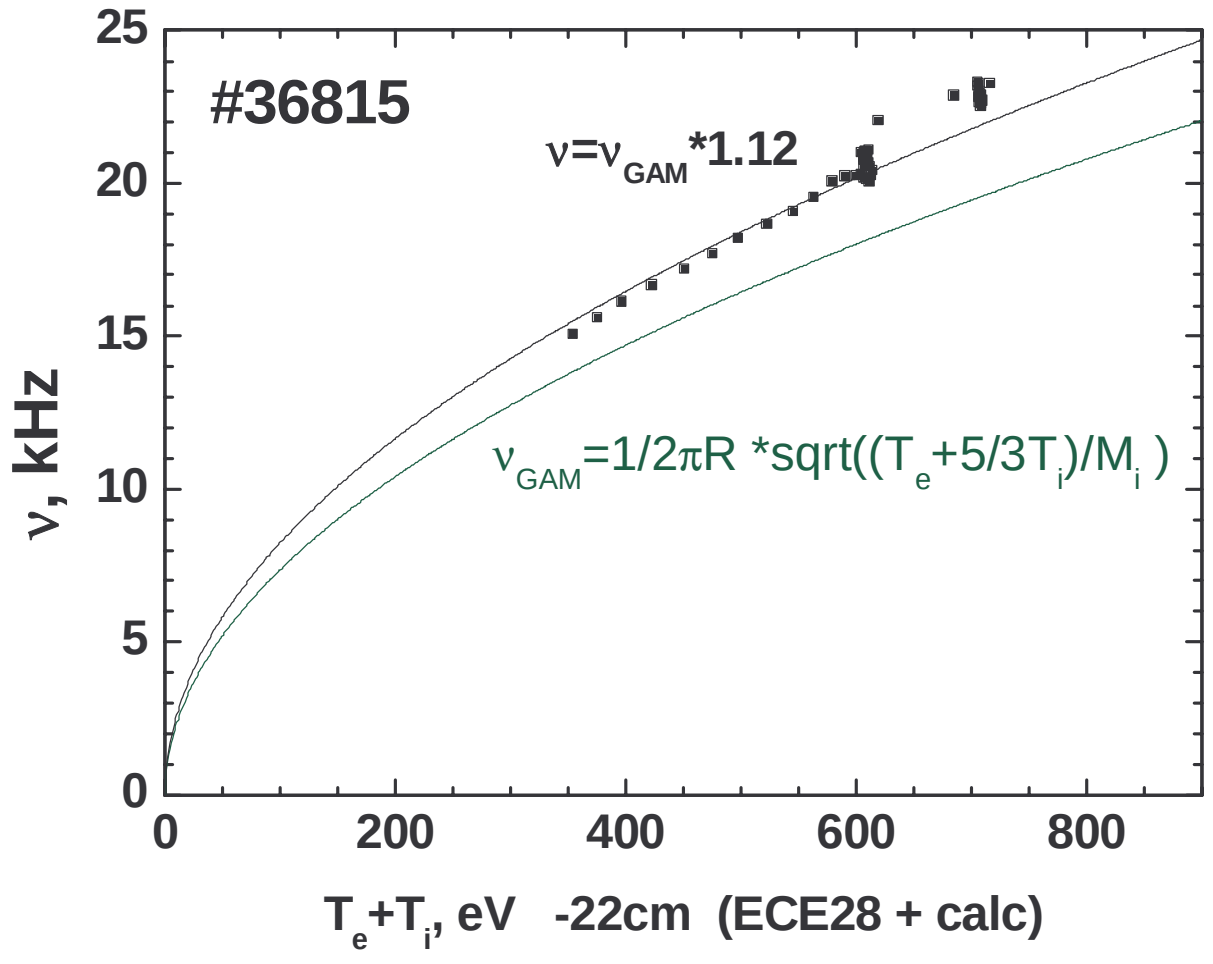

**Figure 10.** Comparison of the “20 kHz” mode frequency with GAM frequency calculated in two-fluid approximation, $\nu_{exp} \sim K\,\nu_{GAM}$, $K$~1.

As it was mentioned, the discussed “20 kHz” mode is seen mainly on the potential signal (Fig. 11, left). In studied regimes its frequency is varied from 9 till 26 kHz, being outside the typical MHD frequency range. On the signal of total beam current (Fig. 11, right), which is proportional to density, this GAM type of oscillations practically absent, but we see oscillations with frequencies about 7 kHz, which are the indication of the MHD activity m=2. We should note that the same case was seen in TEXT [1]: MHD peak dominates in the beam current power spectrum, while “20” kHz peak dominates in the potential power spectrum.

The estimations of the absolute values of the potential and density oscillations give us following:

$A$(25 kHz) ~ 60 V = $\Delta\varphi$ ; $T_e \sim 400$ eV, $e\,\Delta\varphi/T_e = 1.5\times10^{-1}$; $\Delta n/n < 10^{-2}$.

We see that $(e\Delta\varphi/T_e)/(\Delta n/n)>10$, i.e. the Boltzmann relationship is not fulfilled for this kind of the oscillations and they are not driven by density fluctuations.

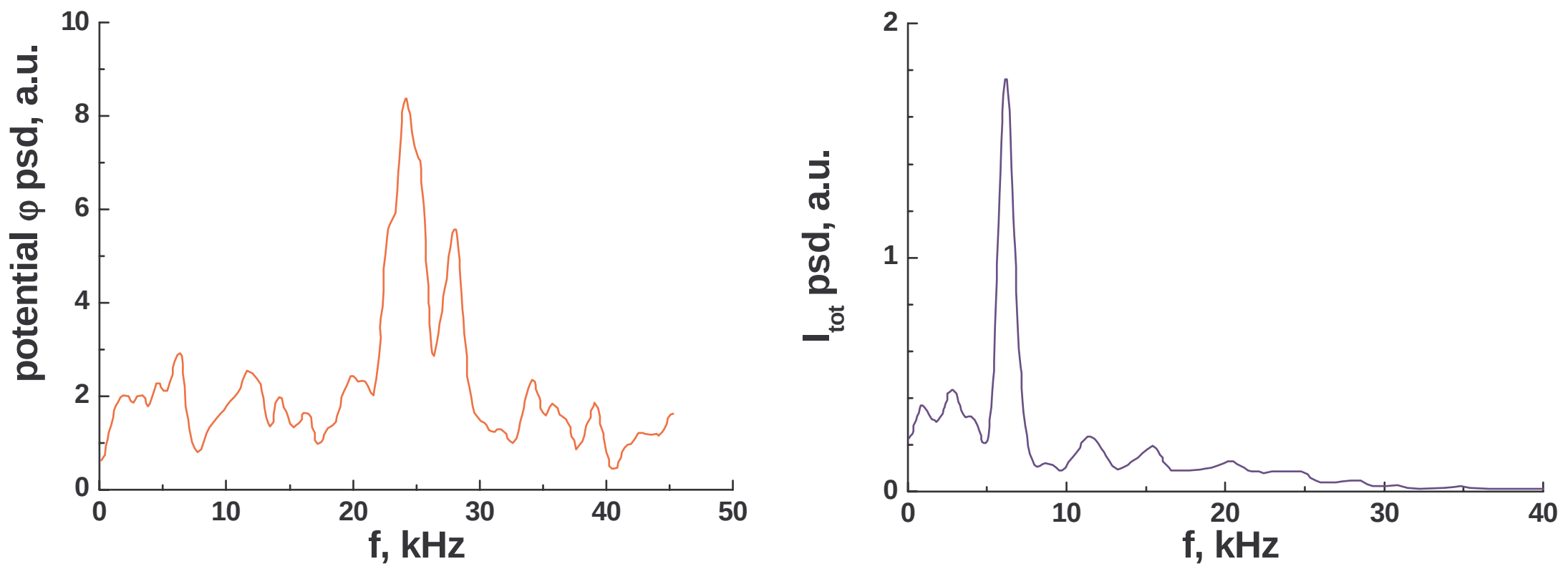

**Figure 11.** Comparison of power density spectra for the plasma potential (left) and the total beam current (right).

Comparison of our observations with HIBP measurements of plasma potential [1] and electric field $E_r$ fluctuations [2] in TEXT, and BES measurements of poloidal rotation velocity $V_{pol}$ in DIII-D [3], shows the significant similarities among the data from all machines. The power spectra of three closely linked parameters: $\Delta\varphi$, $E_r$ and $V_{pol}$ are looking very similar. They have the narrow dominant peak with a high contrast to the noisy background. The absolute values of the frequencies are close to the GAM ones, being slightly smaller. They have the square root dependence on $T_e$.

## 6. Conclusions

The low-frequency potential oscillations were studied in the bulk plasma of the T-10 tokamak by HIBP. The “20 kHz” mode is the dominant high contrast peak on the potential power density spectrum. This kind of plasma oscillations is pronounced mainly on the signal of potential. In studied regimes its frequency is varied from 9 kHz till 26 kHz. On the signal of total current, this type of oscillations practically absents. The “20 kHz” mode frequency depends on the local electron temperature in the sample volume, $\nu \sim T_e^{1/2}$. Observed features are typical for Geodesic Acoustic Mode (GAM). The absolute values of the frequencies are close to the GAM’s one with a factor of an order of unity.

## Acknowledgements

The work is supported by Federal Atomic Energy Agency of RF, and Grants RFBR 02-02-17727, NSh-1608.2003.2, INTAS 2001-2056 and NWO-RFBR 047.009.009.